\documentclass{article}
\usepackage[utf8]{inputenc}
\usepackage[square, numbers]{natbib}
\usepackage{graphicx}
\usepackage{url}
\usepackage{xargs}                      
\usepackage[pdftex,dvipsnames]{xcolor} 
\usepackage[disable]{todonotes} 
\usepackage{xspace}
\usepackage{amsthm}
\usepackage{csquotes}
\usepackage{enumerate}
\usepackage[in]{fullpage}
\newcommandx{\cjc}[2][1=]{\todo[linecolor=blue,backgroundcolor=blue!25,bordercolor=blue,#1]{CJC:#2}}
\newcommandx{\vjt}[2][1=]{\todo[linecolor=green,backgroundcolor=green!25,bordercolor=green,#1]{VJT:#2}}
\newcommandx{\bipr}[2][1=]{\todo[linecolor=red,backgroundcolor=red!25,bordercolor=red,#1]{BIPR:#2}}
\title{Vulnerabilities in the use of similarity tables in combination with pseudonymisation to preserve data privacy in the UK Office for National Statistics' Privacy-Preserving Record Linkage}
\date{\today}
\author{Chris Culnane, Benjamin I.~P. Rubinstein, Vanessa Teague\thanks{School of Computing and Information Systems, University of Melbourne, Australia. \newline Email \texttt{\{cculnane, brubinstein, vjteague\}@unimelb.edu.au}}}
\usepackage{graphicx}

\begin{document}

\maketitle
\section*{Summary}
In the course of a survey of privacy-preserving record linkage, we reviewed the approach taken by the UK Office for National Statistics (ONS) as described in their series of reports ``Beyond 2011". Our review identifies a number of matters of concern. 
Some of the issues discovered are sufficiently severe to present a risk to privacy. The issues discovered are as follows, in order of severity, from least to most severe:
 
\begin{enumerate}
    \item Incorrect cryptographic assumptions have been made, in combination with incorrect statements regarding the required entropy for HMAC keys. The consequence is an overstatement of the security of the solution;
    \item The provision of similarity tables with HMAC'd names exposes the approach to frequency attacks; and
    \item Plaintext similarity scores provide an index to HMAC'd names, permitting plaintext recovery of names.
\end{enumerate}


\section*{Note on Statistical Research Environment}
Our analysis has been performed without access to the underlying data or the Statistical Research Environment (SRE). As such, it reflects on the methods described in the ``Beyond 2011" series only, and specifically focuses on the fact that sole reliance on HMAC'ing in combination with similarity tables is not an appropriate method for protecting data. As noted in the ONS M10 paper \cite{ONSM10}, the SRE itself should provide protection of the data, through physical, security and procedural controls - for example restrictions on what can and cannot be taken into the SRE. The Environment itself should be  fully isolated from all other systems and networks, and should include protective monitoring measures. It is important to note that as such, the underlying data remains protected by the SRE and its corresponding controls – the attacks are not technically difficult, so they could potentially be run within the SRE.  The data should be protected by procedural controls on the basis that it is identifiable. This demonstrates the value of SRE based access to data, in that even when one procedure is shown to contain a flaw (in this case, the use of similarity tables), the other layers of protection allow the underlying data to remain protected. 

Our motivation for raising the issues contained within this report is twofold, firstly, to correct the narrative around the security offered by hashing, HMACs, and similarity tables. This is important as a misunderstanding can propagate more widely and lead to incorrect decisions being taken when protecting data outside of an SRE. Secondly, to correct the incorrect assertion that the data protection methods provided by the HMAC and similarity tables are secure against all currently-feasible attacks and thus can be deployed outside of an SRE to provide protection. As we shall show the methods are susceptible to both existing and new attacks, and should not be deployed outside an SRE. 

We would like to emphasise the attacks contained within this report do not constitute an attack against the SRE itself, they are an attack against a particular set of data protection techniques deployed within the SRE.

We would like to thank the ONS for their constructive response to our analysis.
It is commendable that the ONS has published its methodology, and we hope that our analysis is exactly the sort of constructive participation that should happen between academia and government.  It is a good thing when weaknesses are identified, because then they can be corrected, and the methodology improved before it is used in an inappropriate way.  This process bears witness to openness and transparency being drivers for better security and privacy. Rather than being criticised for making a mistake, the ONS should be commended for making enough details of their methodology public to allow independent scientists to identify weaknesses and have them corrected---this is in the best interests of the privacy of citizens.

\section{Incorrect cryptographic assumptions}
When analysing cryptographic protocols it is essential to be precise about both the components that make up that protocol, and the corresponding assumptions associated with those components. Usage of the correct terminology is absolutely critical to performing an accurate security analysis. Different cryptographic primitives provide different properties under different assumptions. Incorrect usage of terminology can lead to a critical misunderstanding of the cryptographic properties being offered. It is also of equal importance when composing multiple protocols together, to ensure that one protocol/process does not undermine the security of another. 

The M9 document \cite{ONSM9} makes the following statement:

\begin{displayquote}
``We have used a cryptographic hash function to anonymise person identifying information
including names, dates of birth and addresses. The hash function, which converts a field into a
condensed representation of fixed value, is a one-way process that is irreversible – once the
hashing algorithm is applied it is not possible to get back to the original information without
significant effort and the use of tools that are not available in the research environment."
\end{displayquote}

There are two issues with this statement; firstly, cryptographic hashes only approximate one-way functions under certain assumptions, which are not met in the case of the ONS approach. Secondly, the approach does not strictly use a hash function, it uses an HMAC.

\subsection{Hashes and one-way functions}
Most of the security analysis contained within M9 \cite{ONSM9} and M10 \cite{ONSM10} is incorrectly based on the assumption that hashing is a perfect one-way function. However, a cryptographic hash function only approximates a one-way function if the set of pre-images is sufficiently large to prevent brute-force attacks, and if the input is drawn  uniformly at random. In layperson’s terms, this means that the set of all possible inputs is sufficiently large to make it infeasible to try them all, and that all items are equally likely to appear. 

In terms of what would be considered sufficiently large, it should be equivalent to the security level for a symmetric cipher, {\it i.e.} 112 bits of entropy at the very minimum. This equates to $2^{112}$ different possible inputs, that are all equally likely to appear. It should be obvious that none of the ONS entries come even close to this size. As such, a hash function alone would be susceptible to a brute-force attack in what is generally referred to as a dictionary attack. 

A dictionary attack consists of generating hashes of many, if not all, possible inputs and then matching them to the dataset to effectively reverse the contents. In the case of names this is quite trivial, given that the set of possible names is relatively small in cryptographic terms, and is largely considered to be in the public domain. 

\subsection{HMACs vs hashes}
It might seem like distinguishing between an HMAC and a hash is little more than pedantry, but it is not: their security properties are very different. The use of the key in the HMAC prevents dictionary attacks, since an attacker is unable to reconstruct the values without access to the key. It is the entropy and secrecy of the key that is crucial, as opposed to access to software that can construct hashes or HMACs. This is important, since there is a false assumption that the setup in the Statistical Research Environment (SRE) does not provide access to such software. The reality is that access to hashing and HMACs is near ubiquitous, they are provided as part of Microsoft's Cryptographic API, and are included in many software packages including R. As such, the assertion that users of the SRE do not have access to such tools is incorrect. 

The underlying hash function in an HMAC  must provide a combination of collision resistance and near random output. What is critical to the security of an HMAC is the entropy of the key. Both the NIST~\cite{fips2008198} and RFC~2104~\cite{krawczyk1997rfc} specification of the HMAC algorithm make this clear. The key should be generated to the same standard as a symmetric encryption key. The M10 document~\cite{ONSM10} states that: 

\begin{displayquote}
``...provided the keys are handled appropriately, their entropy is not a critical factor in the final hash strength, and thus deterministic pseudorandom number generators could be used..."
\end{displayquote}

This statement is fundamentally incorrect. The security of the HMAC is dependent on the entropy of the key.  If someone re-used this technique, believing the statement to be true, and used a predictable key in their HMAC, then the protection would be no better than an ordinary hash function.  If an attacker could guess the key, then a dictionary attack would be straightforward and result in all names (or other inputs) being easily reversed.

\vjt{I've changed this a bit to be more explicit about why it's bad.  At the risk of channelling Jeff, this language was rather emotive.  I'd leave out ``The quoted statement undermines the credibility of the security analysis.'' It's true, but I think it speaks for itself. }

\subsection{HMACs and frequency}
If the key is generated carefully, the use of an HMAC instead of a hash function protects against the problem of having only a small input set. However, the use of an HMAC does not solve the second problem of having a non-uniform distribution in the input set. HMACs, like hash functions, are deterministic, and as such, the same input will generate the same output when performed with the same key. In the case of the generated matching keys this is not such an issue, since they are largely unique\footnote{the linking keys are still susceptible to frequency attacks in certain situations. For example, where a linking key is non-unique it could present an attack vector for frequency attacks.}. However, the use of an HMAC containing just first or last name in the similarity table is extremely susceptible to frequency attack, and undermines the security of the entire approach. We will discuss this further in the next section. 

\section{Similarity tables at risk of frequency attack}
M10~\cite{ONSM10} contains the statement that

\begin{displayquote}
``...the approach described in this paper is designed to resist all currently-feasible attacks on the basis that the same approach may be used in future (and/or elsewhere) in less controlled environments."
\end{displayquote}

This is not correct. The issue is that the protocol generates HMACs that contain just first or last name. This is essential for the operation of the proposed fuzzy matching protocols. However, this very requirement creates a security weakness, namely, vulnerability to a frequency attack that would break the semantic security of the scheme, and allow plaintext recovery of the HMAC. The simplest example of this is the popularity of the surname ``Smith". It is so overwhelmingly popular that simply analysing the frequency of the HMAC tags would reveal which one represented ``Smith". The same would be true for first name and potentially to some degree for date of birth.

From a security point of view we would not distinguish between the recovery of one cipher text and the recovery of all. A single recovery would be deemed unacceptable. Even if such a recovery was deemed acceptable, the composition of the rest of the protocol results in this single piece of information being leveraged to gain further information, as we shall elaborate on in the next section. 

\section{Similarity scores allow recovery of names}
The provision of a similarity table with plaintext similarity scores critically undermines the privacy of the scheme, facilitating large-scale recovery of names with minimal effort. 

\subsection{Similarity leakage}\label{sec:ons:simleak}
The similarity tables provide plaintext similarity scores, which has the unintended consequence of providing an index to reverse the corresponding names. Since the similarity scores are calculated using a reproducible function, using publicly accessible information (e.g. a list of last names), there is nothing to prevent the scores being recalculated independently outside of the SRE. Once reconstructed it becomes a simple task of matching the plaintext similarity scores to be able to recover the plaintext name. 

By way of an example, we calculated the similarity table for the list of Australian last names (approximately 351,070 names). As described in the supplement to M13~\cite{ONSSuppM13}, a threshold of 25 was used for the SPEDIS\footnote{SPEDIS is an SAS proprietary string comparison method.} scores of first and last name. We applied the same threshold of 25, meaning we discarded any similarity entry with a score over 25. We then sorted the similarity scores into ascending order and compared the uniqueness of each of these sets. If an ordered set of similarity scores is unique amongst other rows in the table it indicates that it can be used as a sort of fingerprint, or index, to the corresponding name in the similarity table available in the SRE. We discovered over 56\% of the sets of similarity scores are unique.
\vjt{I feel that this needs one more sentence of explanation.  I know what you mean, but take is slightly slower and explain that you're trying to re-identify names, that the sorted list of similarities forms a sort of fingerprint, and that those fingerprints are easily calculated and often unique... or something like that ...  or maybe forget about fingerprints and use an example name.  ``As an example, the name ``Smith'' has well-known near neighbours ``Smythe'' (distance 2), ``Saith'' (distance 1), ``Nasmith'' (distance 7) or whatever...  There are unlikely to be any other names with exactly that pattern of distances to the same number of neighbours.}
\cjc{Added an additional line}
As such, by just reconstructing the similarity table it is possible to identify and reverse over half the HMACs. Note: the HMAC key is not required to construct the similarity table, only knowledge of the list of names, which is generally available. Furthermore, in many cases the set of similarity scores that would need to be remembered, in order to act as a unique index, is small enough to be within the capabilities of the average person.

\subsection{Composition of weaknesses}\label{sec:composition}
Even in the situation where an entire similarity table could not be reconstructed, information is still leaked. If we combine the ability to recover a single surname, or first name, via a frequency attack, with the similarity table, we can recover further names. Knowing a single entry, i.e. ``Smith" acts as the start of a chain of recovery of all similar names, which gradually expands to recover many more names. In our example of Australian name data, the ``Smith" entry contains 207 similarity scores, 206 excluding the exact match to ``Smith" itself. Of these 206, 196 are immediately unique by looking at their respective sets of similarity scores. As such, in a single iteration we have recovered 196 names from just knowing which entry relates to ``Smith". The same process can be repeated again for the 196, and again and again for the subsequent similarities in an iterative process. Depending on the threshold selected, and the range of names, it may even be possible to recover the majority of the entries in the similarity table. Again, this does not require knowledge of the HMAC key.

\subsection{Additional Name Entries}
As a result of discussions with the ONS it became apparent that the ONS similarity table contains entries containing common name errors, for example, typographic errors and scanning errors. Such errors could lead to errors like the letter ``B" being replaced by an ``8". This raised the question of whether the addition of these names would make the recovery of plaintexts harder, given that there are a greater number of similarity scores and therefore a greater chance that the set of similarity scores for a particular entry is not unique in the larger ONS set. 

The manner in which the ``error" names are included may allow for their immediate exclusion, for example, if the ``error" names are only included in the right hand column - i.e. incoming names, but not included in the left hand column, a simple set intersection will allow those ``error" names to be excluded and any protection they may provide eliminated. If the ``error" names are included in the left hand column as well this would seem to risk accuracy of matching, in that the best match will be the same error, not necessarily the same individual. Typographic and scanning errors are likely to be transient and not specific and recurring for a particular individual. As such, the errors will match but the underlying records will not. It would seem better for linking accuracy to undertake a cleaning and repair process prior to entry and comparison in the similarity table. Furthermore, if the errors do not appear in an actual dataset, for example, if one dataset contains high quality names (i.e. the census) a set intersection between the HMACs in that dataset and the similarity table will exclude the vast majority of the ``error" names. 

\subsection{Graph Isomorphism}
\begin{figure}[htb]
\centering
\includegraphics[width=0.4\textwidth]{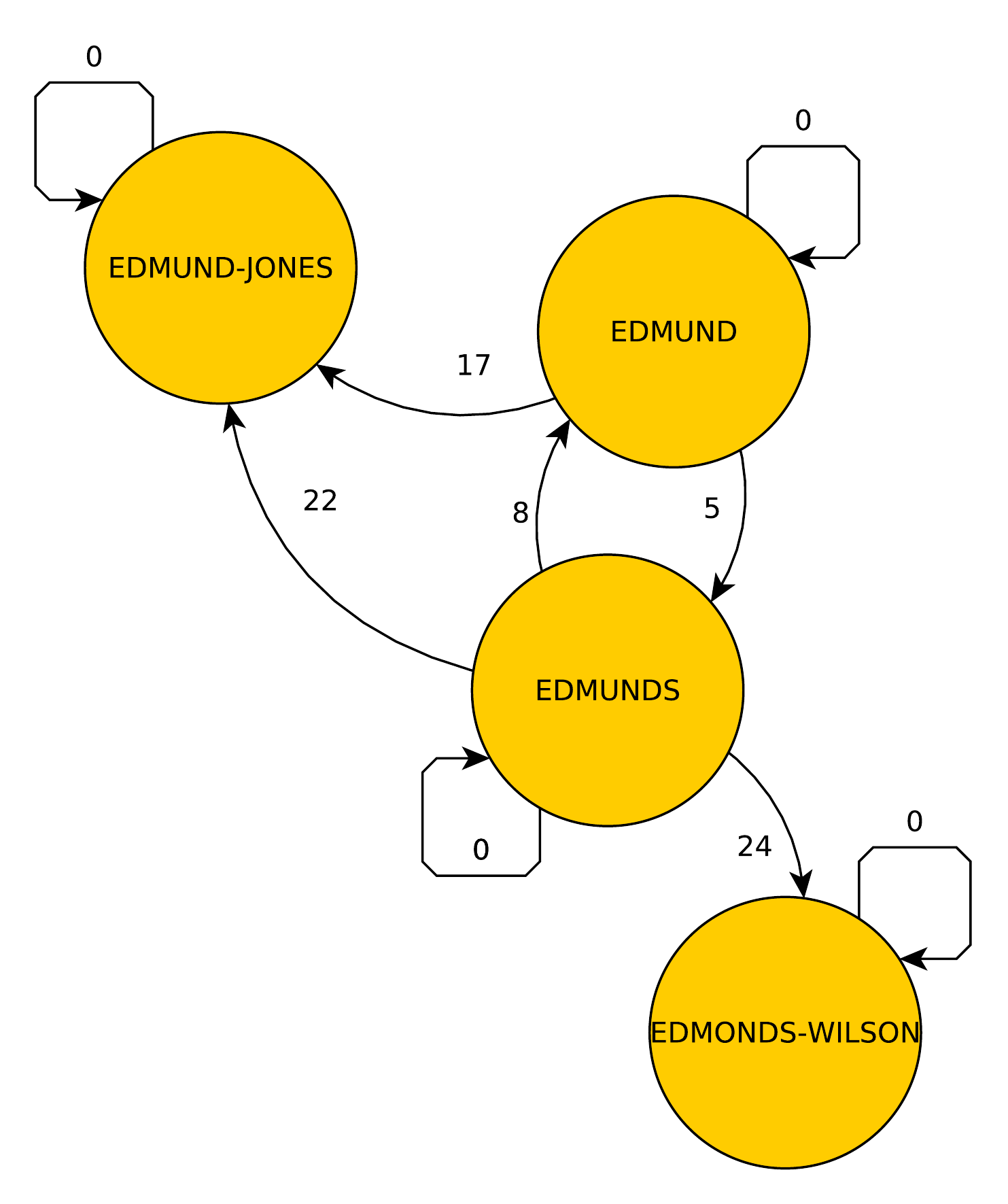}
\caption{Example Similarity Table Graph}
\label{fig:example_graph}
\end{figure}

Putting to one side whether the inclusion of ``error" names is viable, the question of whether only having access to a partial list of names is a valid one, and one which we examine further in this section. In Section \ref{sec:ons:simleak} we took a quick and simple approach to recover plaintext names, we mentioned the notion of chaining together similarity scores, but did not expand on the how to undertake such an action. In essence the similarity table describes a graph in which each node represents an HMAC'd name, and the weighted edges between nodes represent their similarity scores. An example of such a graph is shown in Figure \ref{fig:example_graph}, in this example we have labelled the nodes with the names, but that is just for clarity, in the case of the similarity table those labels would be either blank of the HMAC value.\vjt{This section is great and the pic is really clear.}

\begin{figure}[htb]
\centering
\includegraphics[width=0.9\textwidth]{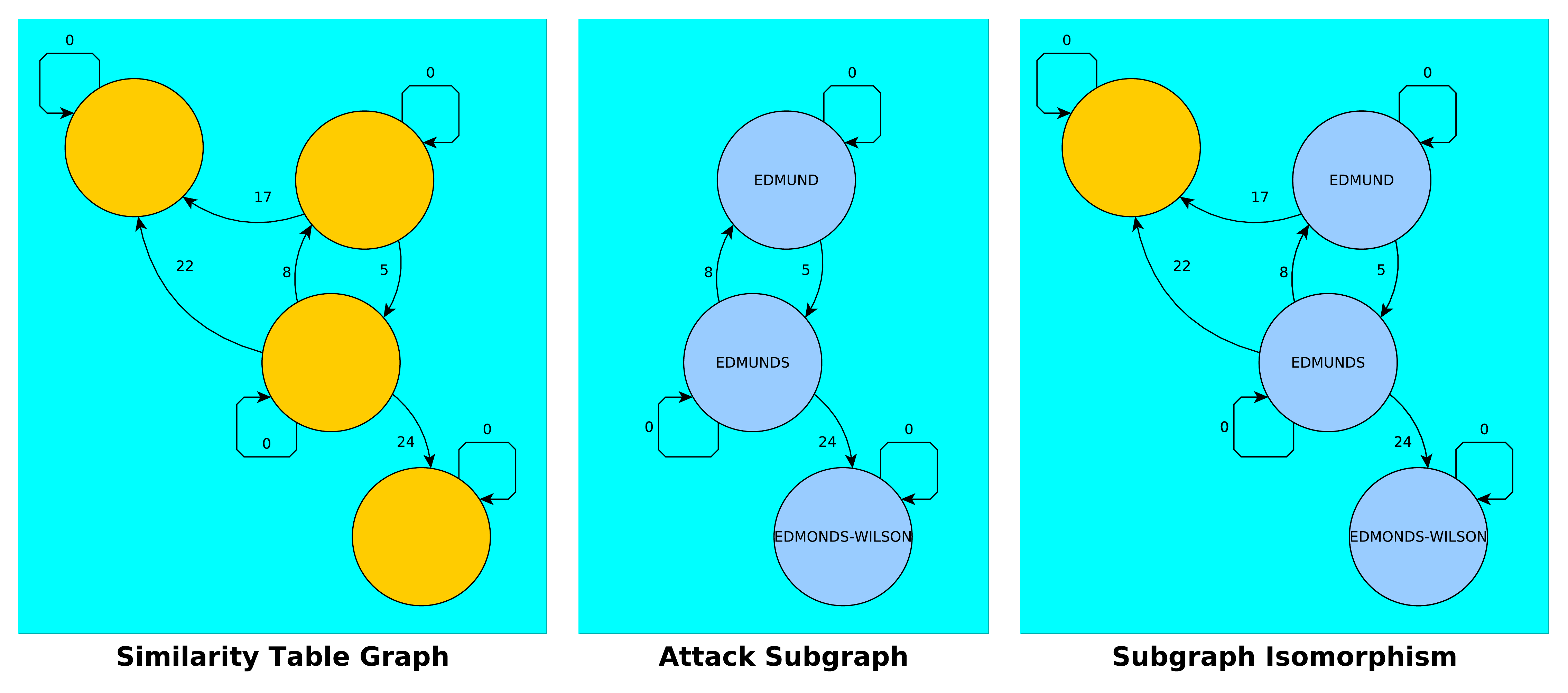}
\caption{Example of Subgraph Isomorphism}
\label{fig:graph_iso}
\end{figure}

The task of recovering plaintexts thus switches from being just finding unique subsets of similarity scores, to a graph isomorphism problem, in particular a subgraph isomorphism problem. In that we can construct a labelled subgraph of names we know to exist and then evaluate whether there is an isomorphism between it and some part of the larger graph constructed from the similarity table. In this way we handle the situation where we do not know all names in the similarity table graph, for example, where ``error" names have been included. The establishing of the isomorphism, based on the edge weights, will allow us to transfer the node label in the subgraph to the similarity table graph and thus recover the plaintexts represented by those nodes. An example of this process is shown in Figure \ref{fig:graph_iso}

In order to simulate having only partial knowledge of the similarity table graph we created a number of samples, of sizes varying from 10\% through to 90\% in increments of 10\%. Each sample was selected independently at random by selecting the respective percentage of names and then only keeping similarity scores for those sampled names. We assume that an attacker can be certain they are constructing a strict subgraph of the ONS similarity table. This is a reasonable assumption, since knowledge of names that are in existence is generally available to the public, {\it i.e.} telephone book.  What is not public knowledge is exactly what additions have been made, but we can be fairly certain names have not been removed. 

\subsubsection{Graph Connectivity}
Due to the SPEDIS similarity score being asymmetric - that is the similarity score between A and B may not be equal to B and A - the graph will also be asymmetric, i.e. there will be nodes that have incoming edges but not outgoing ones. This is shown in our example in Figure \ref{fig:example_graph} for ``EDMUND-JONES" and ``EDMONDS-WILSON", both of which have no outgoing connections other than to themselves. All nodes have a connection to themselves with a similarity score of 0 - to allow exact matching. We have included them in the representation, but they can ignored, since they will never play a part in identifying a specific node, since all nodes have exactly the same zero weighted edge.

We evaluated the number and size of connected components within the various samples tables we had constructed. A connected component is a grouping of nodes that have edges between them. Figure 1 represents a connected component of 4 nodes. When evaluating the connectivity of our different sample sizes it became apparent that the graphs constructed from the similarity scores exhibited similar characteristics. In particular, graphs would have a significant number of singletons (nodes with no connections to other nodes) followed by a small number of groups, not exceeding 102, followed by a single very large connected component containing the vast majority of nodes. Singleton nodes are effectively protected, since singletons are indistinguishable from each other due to them having no defining edges.\vjt{Singletons, of course, are difficult to re-identify because they have no distinguishing edges.  Oh never mind it says that below.  Perhaps put the explanation before the statistics.}\cjc{Moved singleton sentence up to here} At least 93\% of the nodes were accounted for by either being singletons or in a single large connected component. As the size of the sample, and thus the graph grew, the number of singletons rapidly dropped, both in real terms and a proportion of the nodes. Correspondingly, the size of the single largest connected component also grew. 

\begin{figure}[htb]
\centering
\includegraphics[width=0.9\textwidth]{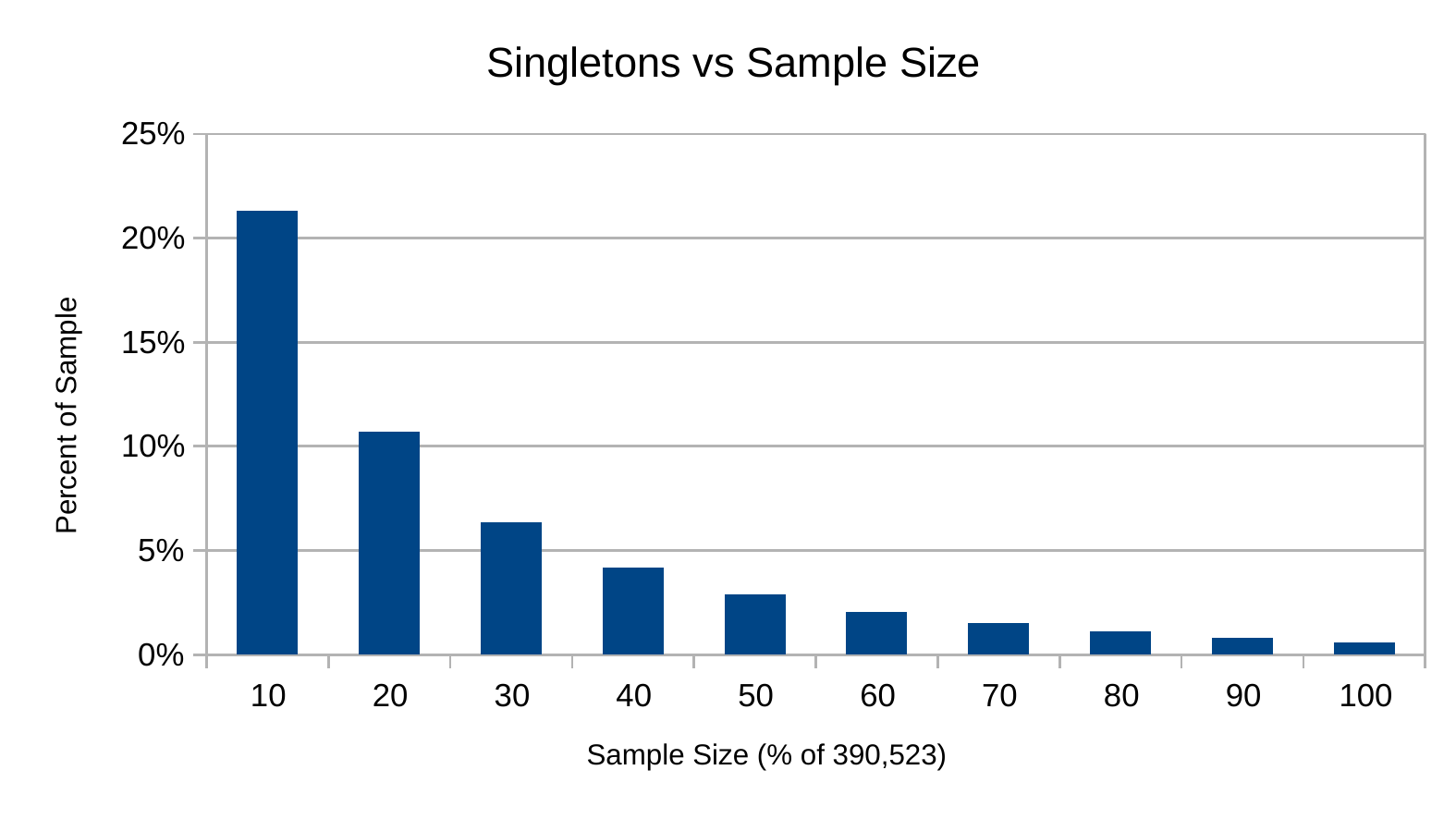}
\caption{Graph of number of singletons vs. sample size}
\label{fig:sample_size}
\end{figure}

Figure \ref{fig:sample_size} shows the rapid decline in percentage of nodes that are singletons as the sample size increases. For a sample size of 10\% (39,052 nodes) just over 21\% of nodes are singletons. Conversely, for a 90\% sample (351,471 nodes) less than than 1\% of nodes are singletons.  

\begin{figure}[htb]
\centering
\includegraphics[width=0.9\textwidth]{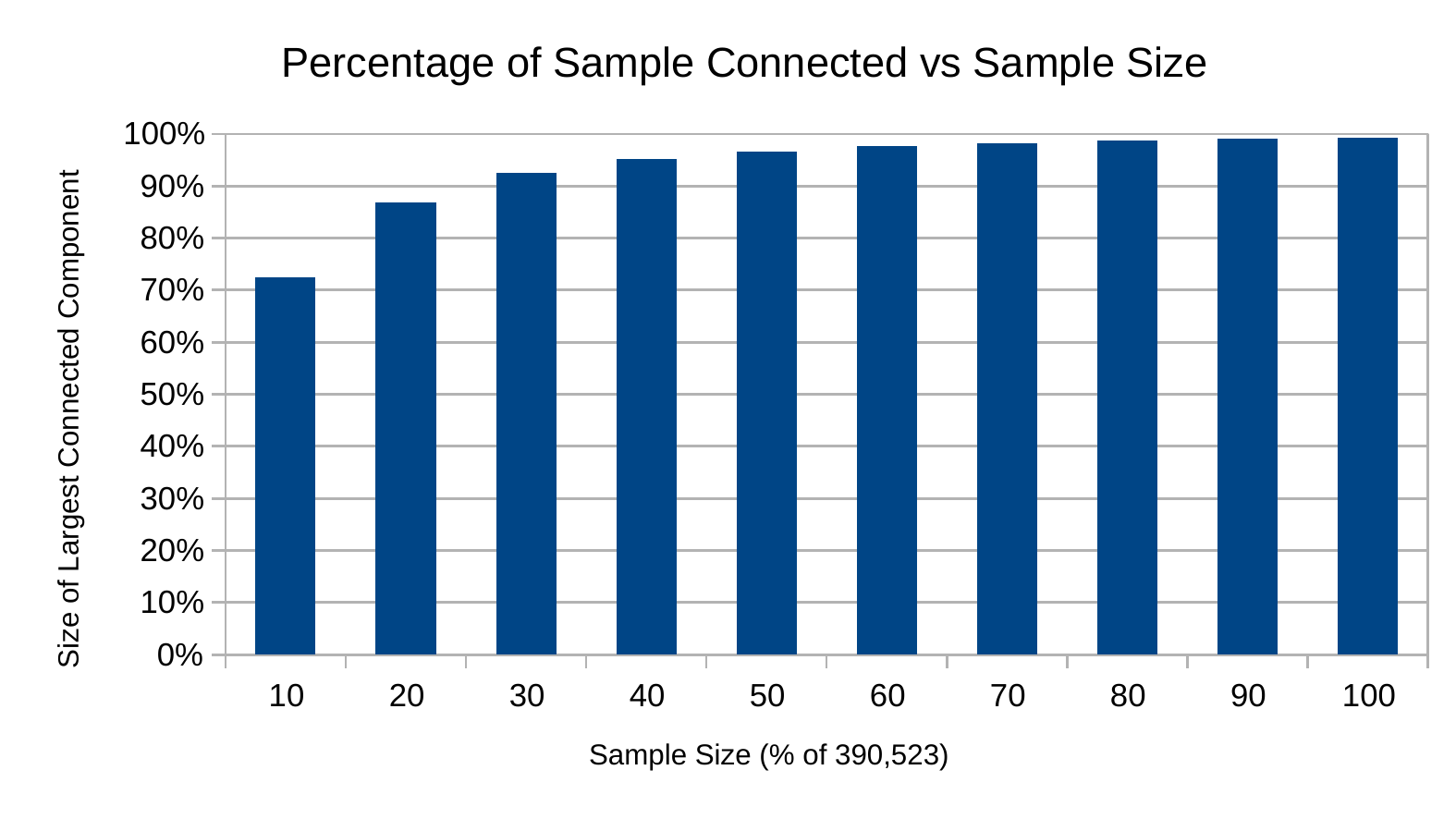}
\caption{Graph of percentage of nodes in a single connected components vs. sample size}
\label{fig:connected_component}
\end{figure}

Correspondingly, Figure \ref{fig:connected_component} shows the percentage of all nodes that are contained in the single largest connected component. We can see that the 10\% sample has a single connected component containing over 72\% of all nodes. At a 90\% sample rate over 99\% of nodes are contained within a single connected component. This level of connectivity is useful for performing the attack. A single very large connected component presents a substantial amount of information, particularly in terms of edge weights, edge direction, and intra-connectivity, providing many features to perform the graph matching on. 

The trend shown by the connectivity results indicates that as the number of nodes grows the level of connectivity increases and the number of singletons fall. As such, it is possible that the addition of additional ``error" names improves connectivity, possibly even providing connectivity to nodes that were previously singletons. As such, when viewed as a graph, the additional nodes may make secrecy of the names worse and aid in performing the plaintext recovery. We can only speculate on this due to not having access to the actual similarity tables, however, the trend is strong in the sampling we have done and as such we would expect the same result to be true for the inclusion of ``error" names. 

\subsubsection{Subgraph Matching}
The process for matching the subgraph to similarity table graph is iterative. The first step is to identify any nodes within the graph that are unique and can obviously be identified. In order to achieve this we first search for any nodes that have a unique set of inbound and outbound edges in the subgraph. By unique we refer to a node having a unique frequency distribution of inbound and outbound edge weights. Referring back to Figure \ref{fig:example_graph}, ``EDMUND" has a frequency of 1 for the inbound weight of 8, a frequency of 1 for the outbound weights of 17 and 5. This is very similar to finding nodes that have unique sets of similarity scores in the similarity table, the only addition is that we now also consider inbound links. 

Once we have found the nodes that are unique in the subgraph we start to look for those nodes in the similarity table graph. We optimise to only looking at those unique in the subgraph because we know if it is not unique in the subgraph it definitely will not be unique in the larger similarity table graph and therefore will not be useful for finding unique nodes in the first pass. For each unique node in the subgraph we check if it is also unique in the larger similarity table graph. If it is unique we consider it to be a match and assign the node from the subgraph to the corresponding node in the similarity table graph. At the same time we adjust the referencing on the incoming and outgoing edges to indicate that this is now a known node - this will be important in the later stages. We continue this for all the unique nodes in the subgraph. If a unique node is found to not be unique in the larger similarity table graph we just ignore it. We do not try and find the set of all possible nodes, since we stop looking as soon as we find a duplicate - this is to optimise the search process to make it more efficient. 

Having labelled all the nodes in the similarity table graph that we can in the first iteration, we move to evaluate each of those newly labelled nodes. For each newly labelled node we compare the outgoing edges, which lead to unlabelled node, with the outgoing edges of corresponding node in our subgraph. If any edge weight is unique for both nodes we can subsequently label the target node of that edge. If there are multiple matching edges with the same weight, we compare the frequency distribution of edge weights for the target node of each edge. In doing this we are checking if there is only a single edge that leads to a target node that provides a sufficient set of edge weights to be a match for corresponding node in the subgraph. If a unique edge is found we subsequently label the target node. Again, we update the inbound and outbound links of the corresponding nodes and queue up the newly labelled node for evaluation. We continue this until we have run out of newly labelled nodes to examine. We then repeat until we complete an iteration without making any changes\footnote{In our experiments this was generally after the first iteration}. 

\subsubsection{Evaluation of accuracy}
Throughout the matching process we only evaluate edge weights and treat the nodes as unlabelled. Having completed our subgraph matching we move to compare all the matches we have assigned for correctness. In order to do this we compare the underlying name for the respective matched nodes in the subgraph and similarity table graph. If they match it is considered a true positive, if they were to not match it would be considered a false positive. In this context we have knowledge of both names since we constructed both graphs. Knowledge of the name in the similarity table graph is only required to evaluate the accuracy of our proposed attack, it is not required to perform the attack itself. 
\vjt{Note that in a real attack this would not be possible---we are able to do this evaluation because we generated both graphs with knowledge of the real names.}\cjc{Added a line about this only being for evaluation and knowledge of names is not required to perform the attack}

Table \ref{tbl:matching} shows the results for the various sample sizes. The percentage of recovered names constitutes the percentage of the subgraph which has been matched. As can be seen, even when we know only 40\% of similarity table we are still recovering 69\% of those names, which exceeds the 56\% we were previously recovering with just the simple set comparison of similarity scores. When we know 90\% of the similarity table we are able to recover over 90\% of the names. As we shall discuss in Section \ref{sec:further_improvements}, the current subgraph matching is far from exhaustive, and as such it is likely we could further improve those results. We have not done so currently because the results for the simple approach overwhelmingly demonstrate the vulnerability of the similarity table.  \vjt{What fraction is this of the non-singleton nodes?}\cjc{I can calculate this, I have it it in the larger table, but didn't think it was that relevant. We don't distinguish between singleton and non-singleton in terms of our desire to recover. The number of singletons as a percentage of the total nodes drops from about 20\% to 1\%}

\begin{table}[htbp]
\centering
{\small
\begin{tabular}{|c|c|c|c|c|c|c|c|}
\hline
\multicolumn{1}{|l|}{\textbf{Sample Size}} & \multicolumn{1}{l|}{\textbf{\shortstack{Subgraph \\ Nodes}}} & \multicolumn{1}{l|}{\textbf{\shortstack{Unique in \\ subgraph}}} & \multicolumn{1}{l|}{\textbf{\shortstack{Unique in \\ similarity \\ graph}}} & \multicolumn{1}{l|}{\textbf{Matches}} & \multicolumn{1}{l|}{\textbf{\shortstack{True \\ Positives}}} & \multicolumn{1}{l|}{\textbf{\shortstack{False \\ Positives}}} & \multicolumn{1}{l|}{\textbf{\shortstack{Percentage \\ Recovered}}} \\ \hline
100\% & 390,523 & 22,381 & 22,381 & 362,934 & 362,934 & 0 & 93\% \\ \hline
90\% & 351,471 & 20,301 & 15,517 & 321,348 & 321,348 & 0 & 91\% \\ \hline
80\% & 312,418 & 19,686 & 10,113 & 279,769 & 279,769 & 0 & 90\% \\ \hline
70\% & 273,366 & 20,680 & 6,057 & 237,402 & 237,402 & 0 & 87\% \\ \hline
60\% & 234,314 & 19,657 & 3,251 & 194,958 & 194,958 & 0 & 83\% \\ \hline
50\% & 195,262 & 17,046 & 1,530 & 151,814 & 151,814 & 0 & 78\% \\ \hline
40\% & 156,209 & 14,047 & 533 & 108,256 & 108,256 & 0 & 69\% \\ \hline
30\% & 117,157 & 11,331 & 136 & 64,391 & 64,391 & 0 & 55\% \\ \hline
20\% & 78,105 & 7,205 & 23 & 24,191 & 24,191 & 0 & 31\% \\ \hline
10\% & 39,052 & 3,079 & 0 & 0 & 0 & 0 & 0\% \\ \hline
\end{tabular}
}
\caption{Subgraph Matching Results}
\label{tbl:matching}
\end{table}

The result for 10\% is zero due to not being able to identify a single node to start the search from. A more thorough subgraph matching could look beyond single node uniqueness to find a starting point, at the cost of additional computation cost. As such, it should not be seen as providing protection from recovery, it just represents a more computationally expensive recovery process, although the percentage of recovered names will be lower, as would be consistent with the trend of reduced recovery as the subgraph size gets smaller. It should be noted that a sample size of 10\% is also unrealistically small for something as commonly known as name, even taking into consideration a noisy set of names. \vjt{Note that 10\% is a very small fraction, probably not at all relevant to the real case of an attacker with a list of known names attempting to re-identify a noisy set.}\cjc{added a comment about that}

Another result of particular interest is that for 20\%. In this result we were only able to identify 23 initial unique nodes to start from. However, the level of connectivity meant that from just those 23 initial nodes we were able to recover 24,191 names in total. This ties back to Section \ref{sec:composition} regarding the risk from the composition of a frequency attack with the similarity table. Currently we do not use frequency information in our analysis, if we did we would be able to identify at least one starting node, and probably many more - for example, identifying the top 50 nodes in terms of frequency and then only looking for unique nodes within that subset. 

\subsubsection{Further Improvements}\label{sec:further_improvements}
Our subgraph matching is currently quite simplistic and far from exhaustive, it is heavily optimised to improve performance and simplify implementation. As such there are a number of steps that could be taken to improve the matching further:

\begin{enumerate}
    \item Implement probabilistic matching, for example, where we have $n$ edges that contain a possible match each one represents a 1 in $n$ chance of being correct. We could further make estimates of node labelling by evaluating the probabilities further.\vjt{Not really relevant for this draft, but I'd suggest combining this idea with the next bullet point.  i.e. if you get multiple matches, explore further down the tree/graph to see which ones are consistent upon further examination.}\cjc{absolutely, I think there is a lot more we could do to improve things further}
    \item Evaluate at further depths where multiple matching edges are found. Currently  we don't explore below a depth of 1, yet we know from the connected component analysis that there is a vast quantity of connectivity information to explore for uniqueness. We have not currently implemented this due to limitations in time and computation resources. 
    \item Evaluate incoming links as well as outgoing edges when trying to find subsequent labels. Currently we only look to label target nodes for outgoing edges for a newly labelled node, we could easily extend this to include incoming edges as well. 
\end{enumerate}

\section{Conclusion and recommendations}
Currently the impact of the weaknesses in the ONS approach is limited by the strict restrictions enforced on accessing the SRE. However, the similarity tables in combination with HMAC'ing remain flawed as a method for preserving privacy of data. If there are insufficient additional controls provided by an SRE, the use of this approach to safeguarding data could lead to the risk of large scale recovery of names. Additionally, the M10 report \cite{ONSM10} states ``...the approach described in this paper is designed to resist all currently-feasible attacks on the basis that the same approach may be used in future (and/or elsewhere) in less controlled environments." In reality, the system's security is derived from the current SRE restrictions and not the cryptographic protocol, and must be considered in the ONS’s current review of protecting data. 

Incorrect usage of cryptographic terms, and a lack of comprehension of cryptographic properties leads to an invalid analysis, and the proposal of a flawed approach mistakenly presented as a secure process. As ONS reports are influential, an important secondary risk is propagation of critical misconceptions by report readers, potentially including other government departments.  

In summary, the use of HMACs in combination with the use of similarity tables is not an appropriate method for preserving the privacy of data outside of an SRE. 

The portion of the approach that uses HMAC linking keys provides reasonable protection, when deployed with suitable assumptions over uniqueness. However, when composed with similarity matching, the scheme as a whole, outside of an SRE, becomes susceptible to a number of practical attacks against privacy. As such we make the following recommendations: 
\begin{enumerate}
    \item The procedural protections around the SRE should be reviewed to take into account that the data is { re-identifiable using public information;}
    \item The incorrect statement regarding the necessary entropy of an HMAC key should be corrected; and
    \item Terminology throughout the reports should be updated to be consistent and accurate, notably, distinguishing between HMACs and hashes, and the associated security analysis updated. 
    \item The similarity table technique should not be used outside of an SRE. 
\end{enumerate}


Again we thank the ONS for their responsible conduct in making their methodological details public to facilitate this kind of analysis.  
Entities that are transparent about their processes and honest about their mistakes, and who work successfully to correct them, deserve more public trust than the rest.  

\vjt{Do we want to say very very clearly that none of this should be given to anyone outside the SRE?}\cjc{Yes, added an additional bullet point}
\newpage
\bibliographystyle{unsrtnat}
\bibliography{references}

\end{document}